\documentclass[a4paper, 12pt, fleqn]{article}

\usepackage[utf8]{inputenc}
\usepackage[T2A]{fontenc}
\usepackage[russian, english]{babel}

\RequirePackage{geometry}
\usepackage[font=footnotesize, labelfont=bf]{caption}

\usepackage[toc,page,titletoc]{appendix}

\usepackage[square,numbers, sort&compress]{natbib}

\usepackage{amsmath,amssymb}

\usepackage[affil-it]{authblk}

\usepackage{graphicx}

\usepackage{multirow}

\usepackage{color}

\title{\textbf{Hubble's Law in Heavy Ion Collisions}}

\author[1]{Kolomoyets~N.V.\thanks{E-mail: \texttt{nkolomoyets@jinr.ru}}}
\author[1,2]{Teryaev~O.V}
\author[1]{Voronyuk~V.}
\affil[1]{Joint Institute for Nuclear Research, Dubna, 141980, Russia}
\affil[2]{Dubna State University, Dubna, 141980, Russia}

\begin{document}
\maketitle

\begin{abstract}
The evolution of the ``microscopic'' Hubble parameter related to the expansion of matter born in heavy-ion collisions was obtained  for nucleons and pions. The calculations were carried out within the parton-hadron-string dynamics~(PHSD) transport model. Au+Au collisions with $\sqrt{s_{NN}} = 7.8$~GeV at~$b~=~2.5,\ 5.0,\ 7.5$, and $10.0$~ fm were considered. A new method for determining the ``microscopic'' Hubble parameter from simulated data was used.
The ballistic motion was obtained for the longitudinal direction after the separation of the nuclei. In earlier times, the evolution of the ``microscopic'' Hubble parameter in this direction was more complicated. 
For transverse directions, an exponential low-time asymptotics of the Hubble parameter was observed. The obtained values of the ``microscopic'' Hubble parameter are about~40 orders of magnitude higher than the cosmological Hubble constant.
\end{abstract}

\section{Introduction}

There is a remarkable interdependence between particle physics and cosmology: particular features of particle interactions direct the evolution of the Universe and determine the features of astrophysical objects~\cite{Lyth:1998xn, Bezrukov:2007ep, Kolomeitsev:2024gek}, while gravitating objects as well as curved space-time as a whole can influence the production and features of particles~\cite{Hawking:1974rv, Ford:2021syk, Parker:1968mv}.
It is noteworthy that the methods used in astrophysics and high-energy physics are sometimes similar: correlations of femtoscopic quantities lead to estimation of the size of particle sources~(see~Ref.\,\cite{Lednicky:1979ig} and references therein), while correlations of interferometer signals allow one to determine the sizes of cosmic radio sources~\cite{HanburyBrown:1954amm}; an analog of cosmic microwave background can be obtained and investigated in collisions of particles~\cite{Mocsy:2011xx, Heinz:2013wva}.

Research over the last decades has shown that the evolution of matter produced in heavy-ion collisions~(HIC), the fireball, resembles the evolution of our Universe~\cite{Floerchinger:2015eta, Heinz:2013wva}. The Big  Bang model makes it possible to explain a lot of present-day features of the observed Universe. Some progress has been made in building a similar model for the ``little bang''~\cite{Heinz:2013wva}; however, its development is still ongoing.

The key parameter for describing the expansion of the Universe is the Hubble constant. It was observed that the velocity of extragalactic objects grows linearly with distance~\cite{Hubble:1929ig},
\begin{equation}
    \label{eq::HL}
    \vec{v} = H\,\vec{r}.
\end{equation}
This is the famous Hubble law and $H$ here is the Hubble constant. Later, it became clear that the parameter~$H$ can change with time and each cosmological solution of Einstein's equations predicts its own evolution of the Hubble parameter as~\cite{Misner:1973prb}
\begin{equation}
    \label{eq:H_def}
    H(t) = \frac{\dot{R}(t)}{R(t)},
\end{equation}
where $R(t)$ is the space scale factor in the metric
\begin{equation}
    ds^2 = dt^2 - R^2(t)\,d\vec{r}\,^2.
\end{equation}
Information on current status of investigations on the cosmological Hubble constant can be found in~Refs.\,\cite{Poggiani:2025hbe, Perivolaropoulos:2021jda}.

An important case of the Hubble motion of matter in space is ballistic motion, that is, motion with velocities constant in time~(scattering of free particles). In this situation, 
\[\vec{v}_a - \vec{v}_b = \frac{1}{t} (\vec{r}_a - \vec{r}_b),\]
which holds for each pair of particles~$a$ and~$b$ in the frame connected with any particle, and the Hubble parameter
\begin{equation}
    \label{eq::ball_scatt}
    H = 1/t
\end{equation}
can be introduced.

The expansion of a fireball also obeys Hubble's law, and one can use a similar parameter~$H$~\cite{Chojnacki:2004ec, Broniowski:2001we, Csanad:2004mm, Baznat:2015eca}. When applied to HIC, it is usually called the ``microscopic'' Hubble constant~\cite{Jipa:2007zz, Besliu:2009zz}. 
The quantity~$R(t)$ in~Eq.\,(\ref{eq:H_def}) in this case is the size of the fireball.
Experimental determination of the ``microscopic'' Hubble constant is not straightforward. Its estimations can be made by comparing experimental data with models~\cite{Csorgo:2002ry} or by using the general similarity of cosmological and ``microscopic'' expansions~\cite{Ristea:2016psq}.

The Hubble-like behavior of a fireball began to attract attention in the early 2000s, starting with experiments at RHIC. The Cracow model~\cite{Broniowski:2001we, Florkowski:2004tn}, which was created to describe the data of those experiments, by construction contains the Hubble-like motion of matter inside a fireball, more specifically, the ballistic motion. 
It is interesting to note that the Hubble-like motion was in fact obtained as early as in 1978~\cite{Bondorf:1978kz} as a solution to nonrelativistic hydrodynamic equations for the isentropic~(with constant entropy) expansion of a nucleonic sphere into a vacuum~(see Eq.\,(3c) in~Ref.\,\cite{Bondorf:1978kz}).
The Cracow model, as well as the relative blast-wave model~\cite{Schnedermann:1993ws, Florkowski:2004tn}, are nowadays used for both immediate description of experimental data~\cite{Schnedermann:1993ws, Broniowski:2001we, STAR:2023uxk, ALICE:2022veq, BMN:2025uqi} and hadron generators~\cite{Amelin:2006qe, Amelin:2007ic, Lokhtin:2008xi, Kisiel:2005hn}. It should be noted that the typical parametrization of transverse velocity in the linear blast-wave model also represents Hubble-like motion. A relativistic hydrodynamic description of Hubble-like motion can be found in~Refs.\,\cite{Chojnacki:2004ec, Chojnacki:2005mp, Csanad:2019lcl}.

Nowadays, interest in HIC at intermediate energies~($\sqrt{s_{NN}}$ around~2 -- 20~GeV) exists. The expectation of catching a critical point in the QCD phase diagram is the main reason for this. There are other important phenomena expected to manifest themselves stronger at these energies~-- ones connected with polarization and collective phenomena~\cite{Lacey:2013qua}. 
The results obtained at RHIC~\cite{STAR:2017ckg, STAR:2022vlo}, GSI~\cite{HADES:2022enx}, and Nuclotron~\cite{BMN:2025uqi} show that we know far not all about matter at achieved conditions.
New facilities, NICA~(Dubna, Russia) and FAIR~(Darmstadt, Germany), are under construction to investigate phenomena in the mentioned energy region.

At the intermediate energies, kinetic models of HIC are 
useful.
They allow~Eq.\,(\ref{eq::HL}) to be applied directly~\cite{Baznat:2015eca, Zinchenko:2022tyg, Inghirami:2021zja, Tsegelnik:2022eoz}. It was found~\cite{Baznat:2015eca, Tsegelnik:2022eoz} that the Hubble-like~(irrotational) motion is the main contributor to the particle flow, while the solenoidal component of the flow is produced by vorticity.

Not only the velocity profile itself, but also its derivatives can be used to investigate Hubble-like motion. In~Ref.\,\cite{Zinchenko:2022tyg}, the divergence of the velocity field was considered as a criterion for the isotropy of the Hubble flow. The expansion rate~(which is a combination of velocity derivatives) was used in~Ref.\,\cite{Inghirami:2021zja} to obtain the evolution of the ``microscopic'' Hubble parameter in central collisions.

In~Ref.\,\cite{Inghirami:2021zja}, the deviation from the ballistic motion in the fireball expansion was observed. A similar deviation was also observed in~Ref.\,\cite{Tsegelnik:2022eoz} for transverse expansion. In our paper, we investigate this issue.

We have investigated the time evolution of the ``microscopic'' Hubble parameter in noncentral collisions separately for nucleons and pions. The latter are the most abundant products of HIC.
Three directions of the fireball expansion are considered separately. The choice of them is standard: the $z$ axis is directed along the beam axis, the $x$ axis is along the impact parameter, and the $y$ axis is perpendicular to the reaction plane. 
The estimation of the Hubble parameter was made using a new method based on the analysis of the statistical distributions of the coordinate derivatives of the velocity field~\cite{Voronyuk:2024ixl}.

Au+Au collisions at $\sqrt{s_{NN}} = 7.8$~GeV and the impact parameters $b = 2.5$, 5.0, 7.5, and 10.0~fm are considered. 
The simulations are carried out within the PHSD transport model~\cite{PHSD, Bratkovskaya:2011wp}. 

The paper is organized as follows. 
Section~2 is devoted to the setup of the~PHSD model.
Some details of simulations are highlighted there as well. 
In Sec.~3 the determination of the ``microscopic'' Hubble parameter is described, in Sec.~4 the obtained results are presented, and the conclusions are made in Sec.~5.

\section{Simulation setup}

The analyzed data are obtained within the parton-hadron-string-dynamics~(PHSD) transport model~\cite{PHSD, Bratkovskaya:2011wp}. In this section, a brief overview of the related features of the model as well as its implementation and setup are made.

The PHSD is a transport approach based on solving the~Kadanoff-Baym equations in the linear approximation on gradients~\cite{Cassing:2008nn, Cassing:2008sv, Cassing:2009vt}. These equations determine the dynamics of the degrees of freedom of the model: partons, hadrons, pre-hadrons, and strings. 
The last two objects are resonances that decay into partons and hadrons~\cite{Cassing:2008rx, Cassing:2009vt}. 
In our research, we are interested in hadrons, specifically, in pions and nucleons.

In our calculations, just nucleon participants and born particles are taken into account. The spectators~(noninteracting nucleons) are separated out using the rapidity variable
\begin{equation}
  y = \frac12 \ln \frac{E + p_z}{E - p_z}
\end{equation}
with the following condition: if for a nucleon $||y| - y_{\text{beam}}| > 0.27$, it is considered as a participant, otherwise as a spectator.

\begin{figure}[h]
\centering
    \includegraphics[width=0.31\textwidth]{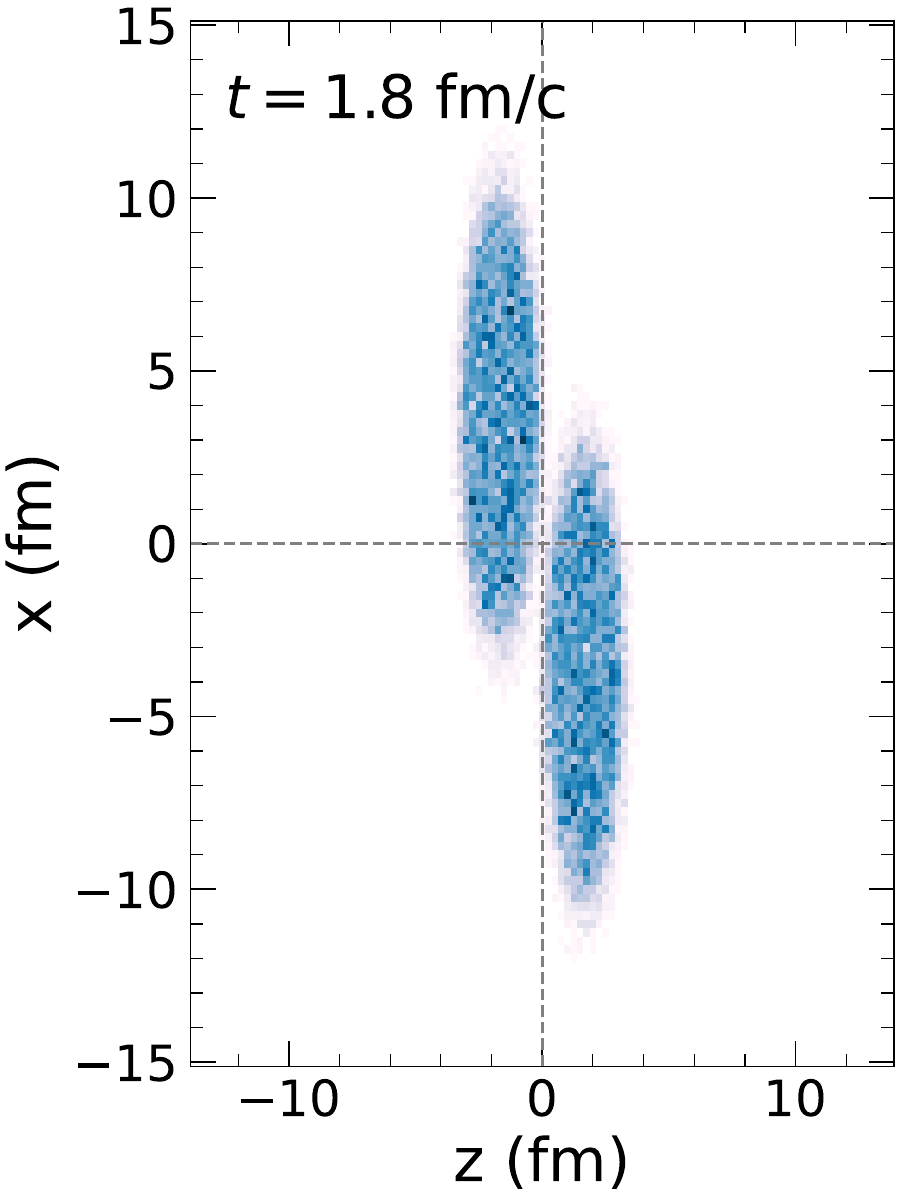} \quad \includegraphics[width=0.31\textwidth]{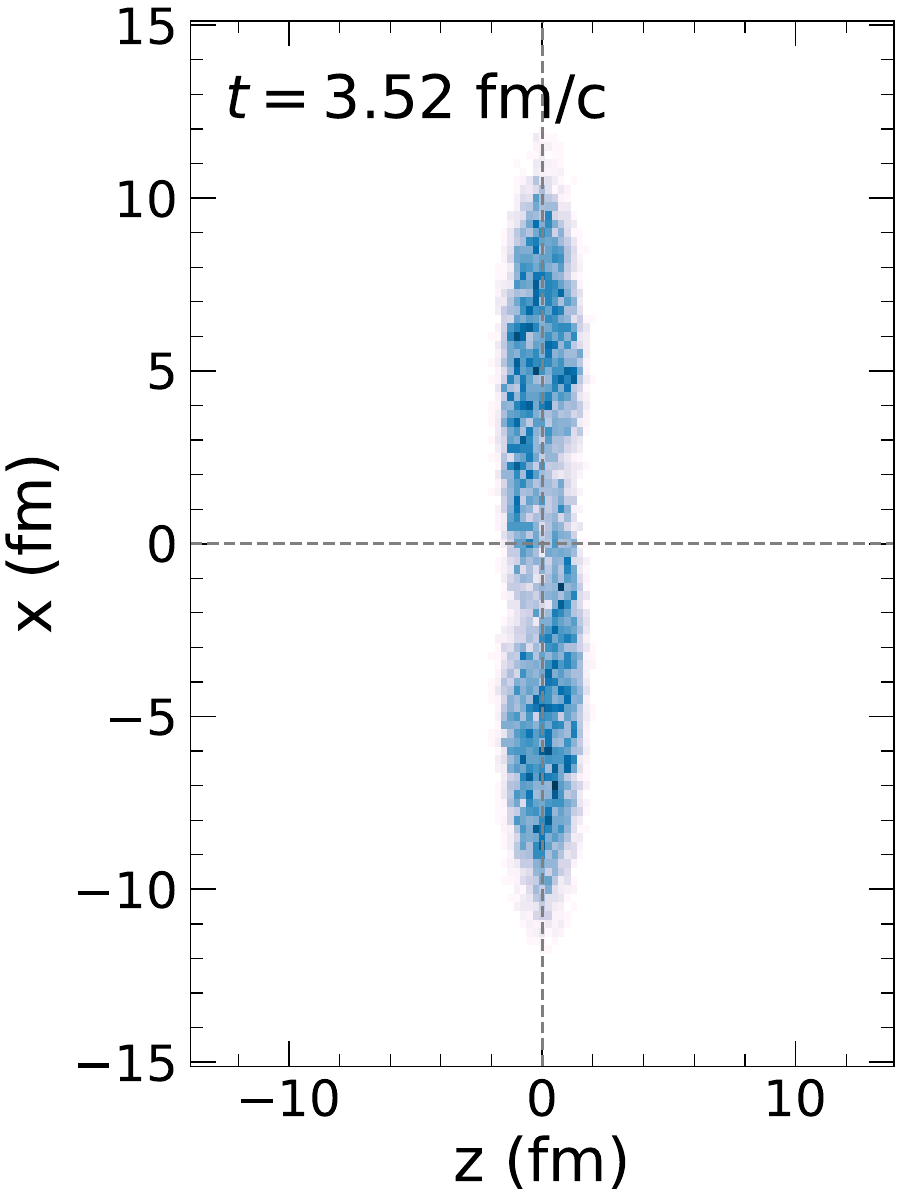} \quad \includegraphics[width=0.31\textwidth]{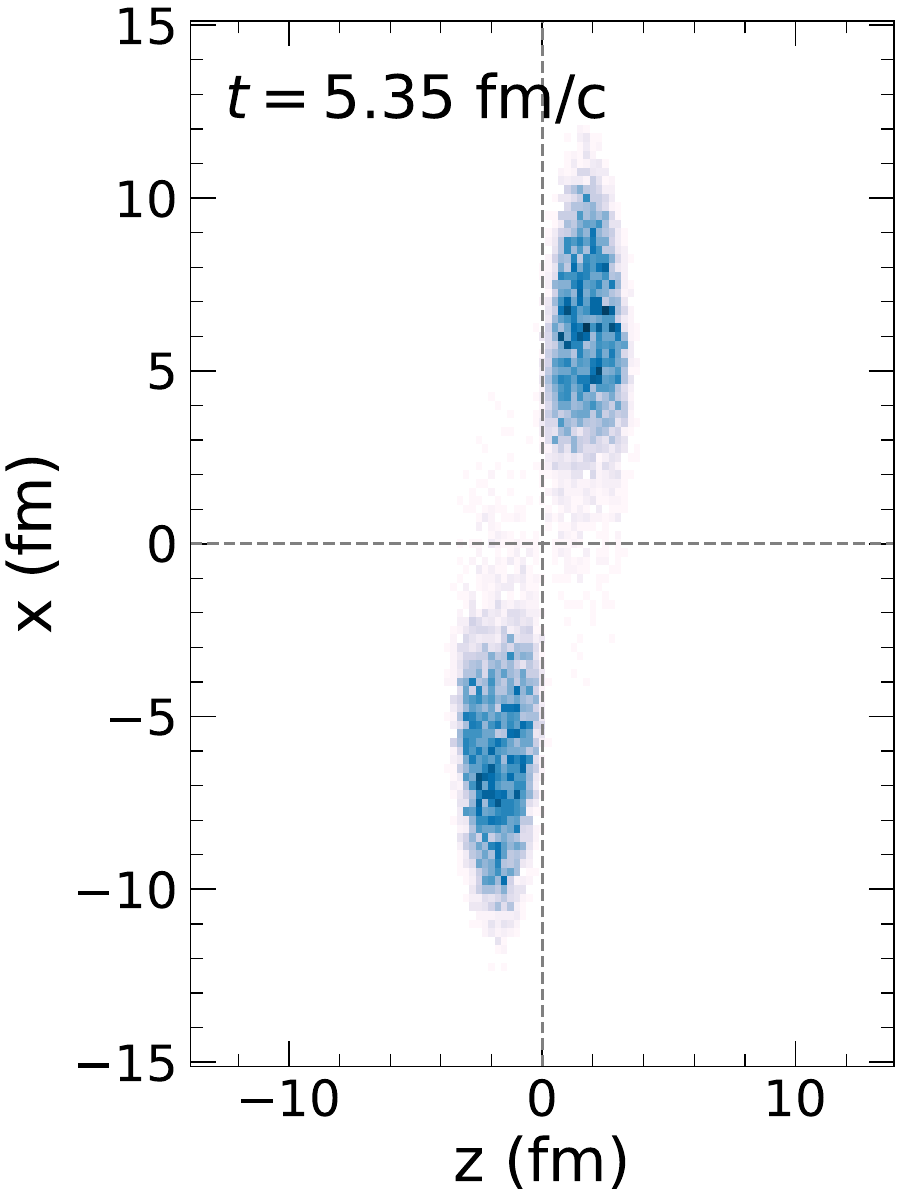}
\caption{Density of spectators projected on the reaction plane~(Au+Au collision at~$\sqrt{s_{NN}} = 7.8$~GeV,~$b = 7.5$~fm).}
\label{fig:SpecMap}
\end{figure}

In the PHSD, at the initial time moment $t = 0$, the nuclei to collide are separated by a distance of
\begin{equation}
    \label{eq::TimeLine_d}
    d = 2 \cdot (1.3\, A^{0.333}/\gamma + 1.5) \mbox{ fm}
\end{equation}
along the $z$ axis. Here $A$ is the mass number of the nucleus, and $\gamma$ is the relativistic gamma factor~(in our case~$\gamma = 4.163$). 
Their velocities correspond to the given~$\sqrt{s_{NN}}$,
\begin{equation}
    \label{eq::v_nucl}
    v_{\text{nucl}} = \sqrt{1-\frac{4 m_N^2}{(\sqrt{s_{NN}})^2}},
\end{equation}
where $m_N = (m_p + m_n)/2 = 0.938$~GeV is the mass of the nucleon. Assuming that the nucleons do not interact with each other inside the nuclei, some key time moments of the collision can be estimated:
\begin{equation}
    \label{eq:tMoments}
    \begin{array}{l}
        \text{(i) the time moment of the maximum overlap: } t_{\text{X}} = (d/2) / v_{\text{nucl}};\\[0.75em]
        \text{(ii) the first touch time moment: } t_{\text{FT}} = t_{\text{X}} - R/(\gamma v_{\text{nucl}});\\[0.75em]
        \text{(iii) the last touch time moment: } t_{\text{LT}} = t_{\text{X}} + R/(\gamma v_{\text{nucl}}),
    \end{array}
\end{equation}
where $R$ is the average radius of the nucleus, 
\begin{equation}
    R = 1.123\, A^{1/3} - 0.941\, A^{-1/3}.
\end{equation}
For Au+Au collisions at~$\sqrt{s_{NN}} = 7.8$~GeV, the estimated $t_{\text{FT}} = 1.84$~fm/$c$, $t_{\text{X}} = 3.41$~fm/$c$ and $t_{\text{LT}} = 4.99$~fm/$c$.
Looking at the simulated spatial distributions of the spectators~(Fig.\,\ref{fig:SpecMap}), one can see that the actual values of $t_{\text{X}}$ and $t_{\text{LT}}$ are somewhat larger, while $t_{\text{FT}}$ is consistent with the result of Eq.\,(\ref{eq:tMoments}):
\begin{equation}
    \label{eq::tMomentsReal}
    t_{\text{FT}} \approx 1.80~\text{fm}/c,\qquad t_{\text{X}} \approx 3.52~\text{fm}/c,\qquad t_{\text{LT}} \approx 5.35~\text{fm}/c.
\end{equation}
The difference is that the radius of the nucleus is somewhat larger than the average one given in Eq.(9) and that the errors are due to the finite time step.
We use the estimates from~Eq.\,(\ref{eq::tMomentsReal}) as reference time moments.

For our goal, we consider matter as a continuous medium. 
For this, the ``fluidization'' procedure~(similar to hydro) is used. 
This means that the space occupied by the fireball is discretized using a lattice, and the basic units of matter are represented by cells of this lattice rather than by separate particles~(see~Ref.\,\cite{Tsegelnik:2022eoz} for more details). The sizes of each cell of the lattice are $\Delta x = \Delta y = \gamma \Delta z = 1$~fm. The medium is assumed \textit{existing} in a cell if the energy density~$\varepsilon > \varepsilon_0 = 0.05$~GeV/fm$^3$ is created by particles inside the cell. We apply this condition separately for nucleon and pion components of the matter. This kind of setup leads to the appearance of the nucleon and pion media at time around~2~fm/$c$.

In this representation, the velocity field defined on the lattice is considered instead of the velocities of separate particles. There are several possibilities to define velocity connected with a cell. The Eckart definition of the velocity~\cite{Eckart:1940te, Deng:2016gyh} is used in this research,
\begin{equation}
    \vec{v}(\vec{x}, t) = \vec{J}(\vec{x}, t) / J^0(\vec{x}, t),
\end{equation}
where~$J^\mu$ is the 4-vector of a particle 
flow. For non-point-like particles it is defined as follows:
\begin{equation}
    \label{eq::Jmu}
    J^\mu(x) = \frac{1}{\mathcal{N}} \sum_a \frac{p_a^\mu}{p_a^0}\, \Phi(x, x_a), \qquad \mathcal{N} = \int d^3\vec{x}\, \Phi(x, x_a),
\end{equation}
the sums are taken over all particles. The smearing function $\Phi(x, x_a)$ is related to the size of particles and defines the transition from discrete quantities related to the $a$th particle with coordinates~$x_a$ to continuous fields of these quantities. In our case, the coordinates~$x$ represent the centers of the lattice cells.
To have smooth distributions, a quadratic spline function is taken as the smearing function~$\Phi(x, x_a)$, the same as in~Ref.\,\cite{Tsegelnik:2022eoz}.

As applied to our research, the PHSD generates an ensemble of velocity-field configurations corresponding to some fixed time moment. There are significant statistical fluctuations of velocities inside one configuration. To suppress them, the values of velocity in each cell are averaged over some number of configurations~($\sim 10^6$). At the edges of the fireball, typically not all the generated configurations contribute to a particular cell. In this situation, averaging is performed over actually contributing configurations. In this way, the field of velocities averaged over configurations is
\begin{equation}
    \label{eq::v_avg}
    \langle \vec{v}(\vec{x}, t) \rangle = \frac{1}{n} \sum_{i = 1}^n \vec{v}_i(\vec{x}, t),
\end{equation}
where $n$ is the amount of configurations that actually contribute
to the cell.
If less than~5\% of the generated configurations contribute to a cell, that cell is excluded from the consideration.

To suppress fluctuations, the method of parallel ensembles is used in the~PHSD~\cite{Xu:2017pna}. For our task this means that a set~(ensemble) of the field configurations is processed simultaneously, and averaging with~Eq.\,(\ref{eq::v_avg}) is applied to them automatically.
However, the amount of parallel configurations is limited by computational resources. We have implemented cell-wise averaging over all sequentially produced ensembles of configurations. For the sake of processing speed up, the OpenCL parallelization of the code was performed.

\section{Determination of the Hubble parameter}

Having the velocity field determined with~Eq.\,(\ref{eq::v_avg}) at time moment~$t$, one can estimate the ``microscopic'' Hubble parameter at that time moment. For this purpose, the profiles of velocity components are typically used, and the Hubble parameter is estimated as a slope of the linear part of that profile~\cite{Baznat:2015eca, Zinchenko:2022tyg, Tsegelnik:2022eoz}. 

Here we use another approach consisting of the analysis of statistical distributions of the directional derivatives of the velocity components~\cite{Voronyuk:2024ixl}.
It is based on transformation of the manifold of the velocity profiles contained in the whole fireball into statistical distribution of the derivatives
\begin{equation}
    \label{eq::part_vi}
    \partial_i v_i = \frac{\partial\, \langle v_i(\vec{x}, t) \rangle}{\partial x_i}, \qquad i \in \{x, y, z\}.
\end{equation}
With the proper normalization, this is the probability-density distribution of the quantity~$\partial_i v_i$ inside the fireball. The position of the nontrivial peak in that distribution corresponds to the average Hubble parameter in the fireball.

The derivatives in~Eq.\,(\ref{eq::part_vi}) are the terms of the divergence~$\operatorname{div} \langle \vec{v}(\vec{x}, t) \rangle$ of the velocity field. The connection of this divergence with the ``microscopic'' Hubble parameter was discussed in~Refs.\,\cite{Baznat:2015eca, Zinchenko:2022tyg}.

\begin{figure}[h]
\centering
    \includegraphics[width=0.318\textwidth]{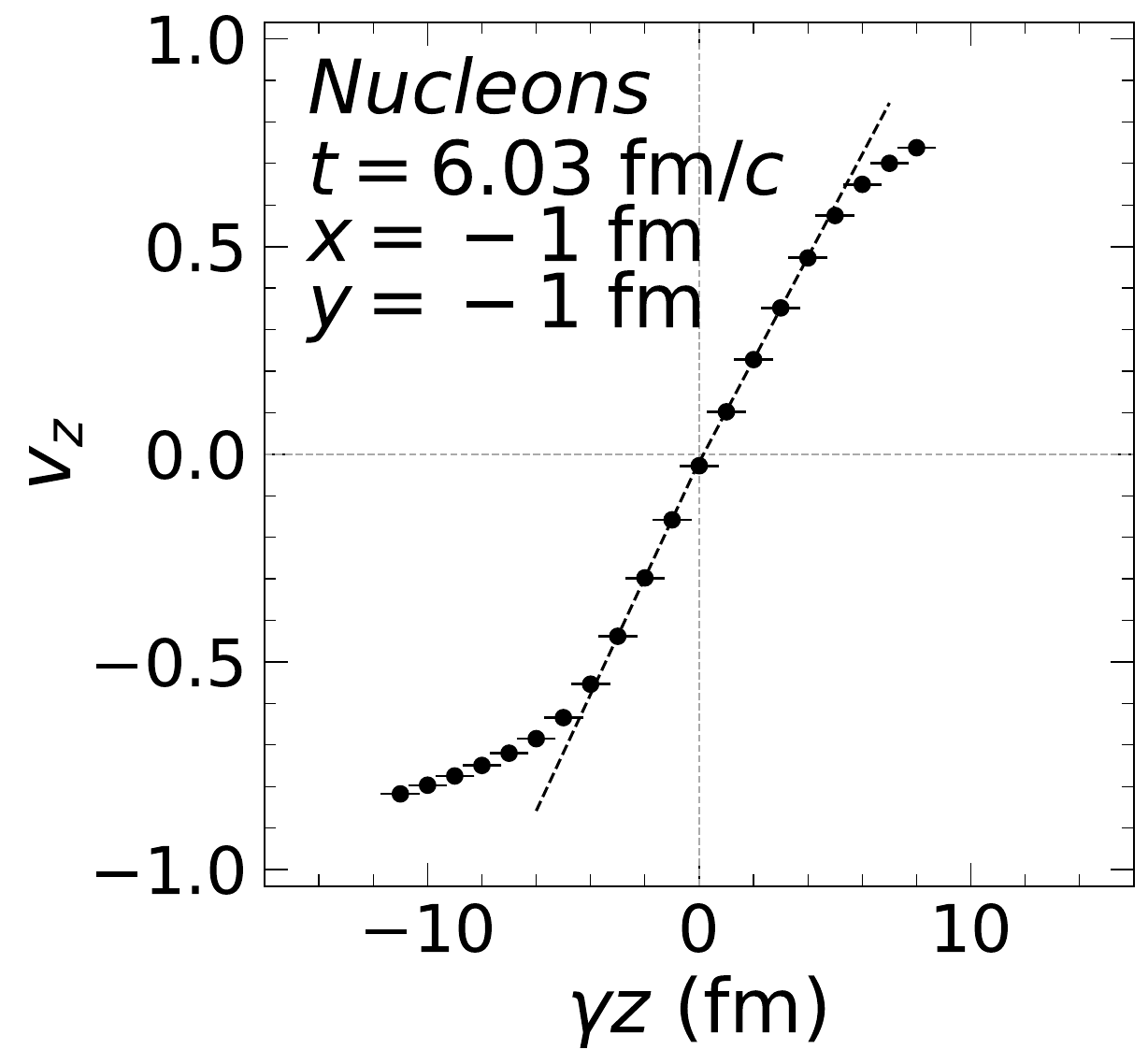}
    \includegraphics[width=0.673\textwidth]{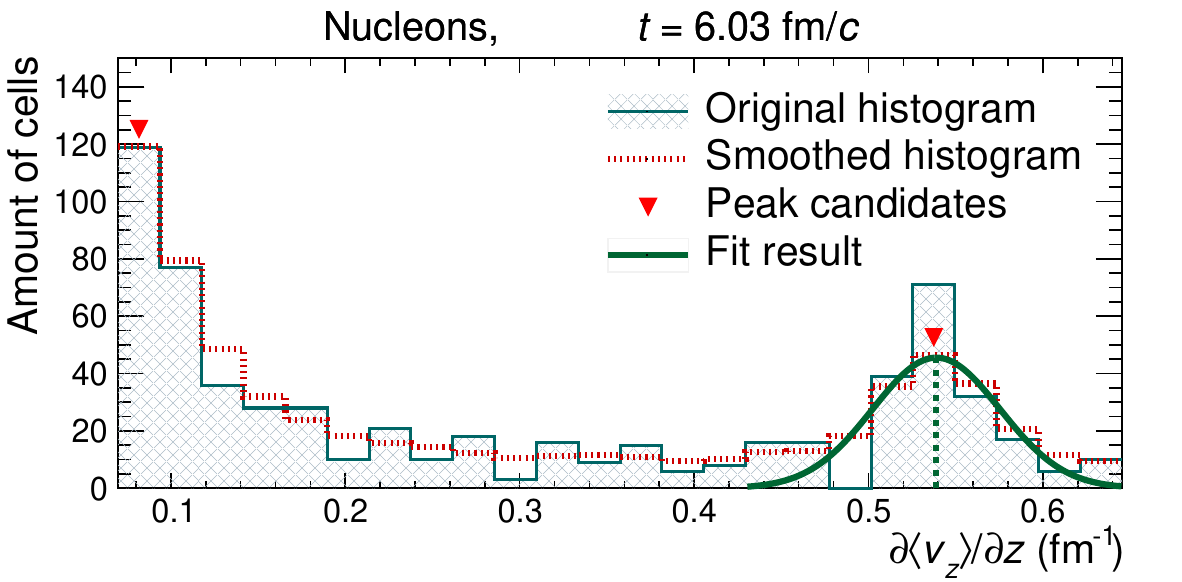}
\caption{A longitudinal velocity profile and the corresponding statistical distribution of~$\partial_z v_z$~(Au+Au collision at~$\sqrt{s_{NN}} = 7.8$~GeV,~$b = 7.5$~fm).}
\label{fig:prof_vs_hist}
\end{figure}

For the velocity profiles appearing in the expanding fireball, 
the statistical distribution of the derivatives from~Eq.\,(\ref{eq::part_vi}) in the general case consists of two peaks: one corresponds to Hubble-like motion and the other emerges only after some time and corresponds to the nonlinear regions at the edges of the velocity profile.
A typical velocity profile and the corresponding histogram of the statistical distribution are shown in~Fig.\,\ref{fig:prof_vs_hist}.

The task comes down to estimation of the position of the peak corresponding to Hubble-like motion. The low~$\partial_i v_i$ peak, when it exists, is mainly pronounced for the~$\partial_z v_z$ derivative. We exclude cells forming the low~$\partial_z v_z$ peak from consideration. Then the position of the needed peak is obtained from Gaussian fit of the remaining histogram. The systematics connected with exclusion of the low~$\partial_z v_z$ peak, as well as with binning the histogram, are taken into account.

\section{Results}

\subsection{Longitudinal expansion}

\begin{figure}[h]
    \centering
    \includegraphics[width = 0.65\textwidth]{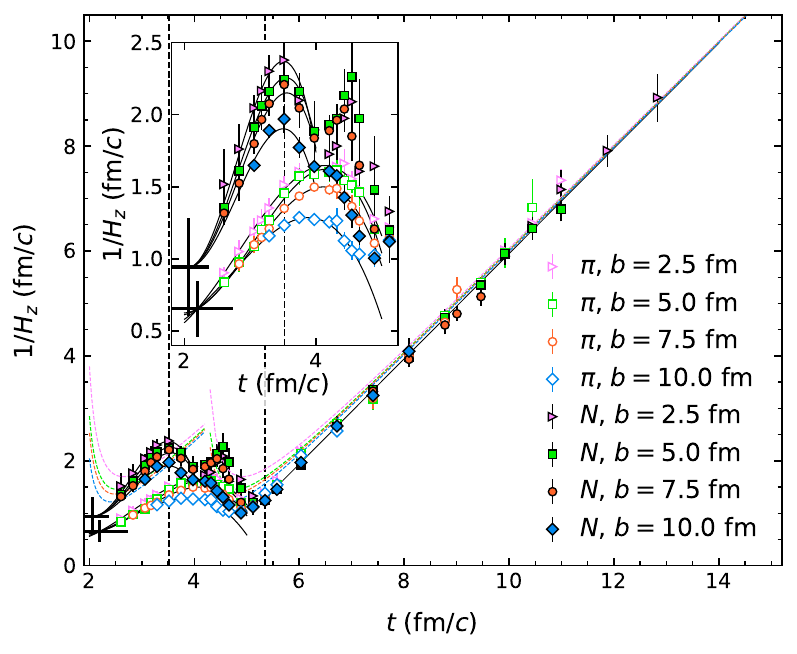}
    \caption{The inverse Hubble parameter in the longitudinal direction. The solid straight line shows the linear fit; the solid curves correspond to polynomial fit given by~Eq.\,(\ref{eq::cube}); the dashed curves correspond to calculation by~Eq.\,(\ref{eq::H_Bondorf}). Two thick crosses show the estimated positions of the intersection points of the fit curves for pions and nucleons. Two vertical dashed lines are for the maximal overlapping moment of time and for the last touch moment. The inset subplot shows zoomed evolution before the linear growth.}
    \label{fig::Hz}
\end{figure}

In Fig.\,\ref{fig::Hz} the evolution of the inverse ``microscopic'' Hubble parameter in the longitudinal direction is shown. One can clearly observe its linear growth after the moment of the last touch of the nuclei. This behavior is common for both nucleons and pions at all considered impact parameters. The common linear fit in that region gives
\begin{equation}
    \label{eq::Hz_bs}
    H_z = \frac{C}{t - t_0},
\end{equation}
with $C = 1.00(1)$ and $t_0 = 4.07(4)$~fm/$c$. 
In this way, the ballistic motion~[see~Eq.\,(\ref{eq::ball_scatt})] with shifted initial time takes place here.
This statement is in agreement with the results obtained in~Ref.\,\cite{Tsegelnik:2022eoz}~(there, no distinction on species of particles is considered) as well as in~Ref.\,\cite{Inghirami:2021zja}~(no distinction on particle species, central collisions, just central cell evolution). 
The estimated value of~$t_0$ is close to the time moment when the maximum particle yields are reached~(in our setup it is around 4.2~fm/$c$).

Before the separation of the nuclei, some transition period is observed. The Hubble parameter here is different for nucleons and pions, and has a weak dependence on the impact parameter. 

For pions, the shape of the inverse ``microscopic'' Hubble parameter evolution is smoother and can be described by the third-order polynomial. 
Independent fits for each impact parameter with the most general cubic function show that all~(extrapolated) resulting curves intersect 
at~$t\approx1.7-2.55$~fm/$c$. It is close to the moment when the fireball emerges. A fit combined for all impact parameters with the function
\begin{equation}
    \label{eq::cube}
    1/H_z^{(\pi)} = 1/H_1 + A_1 (t - t_1) + A_2 (t - t_1)^2 + A_3 (t - t_1)^3
\end{equation}
with a common point $(t_1, 1/H_1)$ gives $t_1 = 2.2(5)$~fm/$c$ and~$1/H_1 = 0.65(18)$~fm/$c$. The estimates for the parameters of~Eq.\,(\ref{eq::cube}) are listed in Table~\ref{tab::Ai}. One can see that both $A_1$ and $A_2$ are compatible with zero. For further improvement of parametrization, the uncertainties~(including systematic ones) should be reduced. Also, estimations of the ``microscopic'' Hubble parameter for earlier time moments are needed. This is not accessible with the current PHSD setup.

We also perform the same combined fit for the first peak of the inverse Hubble-parameter evolution for nucleons. It shows that the linear term in~Eq.\,(\ref{eq::cube}) for this case is essentially zero. The estimates for the rest of the parameters of~Eq.\,(\ref{eq::cube}) with the dropped-out linear term are listed in~Table~\ref{tab::Ai}.

\begin{table}[h]
\centering
\caption{Estimated parameters of Eq.\,(\ref{eq::cube}).}
\label{tab::Ai}
\begin{tabular}{c|c|c|c|c|c|c}
\hline
& \rule{0pt}{1.1em}$b$, fm & $A_1$ & $A_2$, (fm/$c$)$^{-1}$ & $A_3$, (fm/$c$)$^{-2}$ & $t_1$, fm/$c$ & $(1/H_1)$, fm/$c$\\
\hline
\multirow{4}{*}{Pions}&\rule{0pt}{1.1em}2.5 & $0.53(56)$ & $0.25(42)$ & $-0.13(5)$ & \multirow{4}{*}{$2.2(5)$}& \multirow{4}{*}{$0.65(18)$}\\
& 5.0 & $0.25(46)$ & $0.50(27)$ & $-0.19(3)$ &&\\
& 7.5 & $0.29(39)$ & $0.39(26)$ & $-0.16(3)$ &&\\
& 10.0 & $0.41(32)$ & $0.18(29)$ & $-0.12(5)$ &&\\
\hline
\multirow{4}{*}{Nucleons}&\rule{0pt}{1.1em}2.5 & \multirow{4}{*}{---} & $2.07(40)$ & $-0.96(36)$ & \multirow{4}{*}{$2.1(3)$}& \multirow{4}{*}{$0.94(33)$}\\
& 5.0 & & $1.79(31)$ & $-0.81(29)$ &&\\
& 7.5 & & $1.63(23)$ & $-0.73(23)$ &&\\
& 10.0 & & $1.39(17)$ & $-0.65(19)$ &&\\
\hline
\end{tabular}
\end{table}

The fit results for both pions and nucleons are shown in~Fig.\,\ref{fig::Hz} by the solid curves.
Two thick crosses correspond to the estimated intersection points of the fit lines for these types of particles. The sizes of the crosses represent the estimated uncertainties. There is an indication that evolutions of the ``microscopic'' Hubble parameter for pions and nucleons start from the same point. However, additional investigation of the early-time behavior of the ``microscopic'' Hubble parameter should be carried out.

Our results are compared with the results of the nonrelativistic hydrodynamical description. 
From~Ref.\,\cite{Bondorf:1978kz} it follows that for the case considered there,
\begin{equation}
    \label{eq::H_Bondorf}
    H(t) = \frac{t}{t^2 + t_B^2},
\end{equation}
where the characteristic time $t_B$ is given by
\begin{equation}
    \label{eq::t_B}
    t_B^2 = \frac14 \cdot \frac{3 m_N}{\alpha + 5/2} \cdot \frac{n_p + n_t}{n_p\, e_{\text{lab}}} \left[ \frac{(n_p + n_t) m_N}{2\pi \cdot \rho(0) \cdot B(3/2; \alpha + 1)} \right]^{2/3}.
\end{equation}
It is determined by the following parameters: $e_{\text{lab}}$ is the total energy per nucleon in the laboratory frame~(equivalent to fixed-target mode); $n_p$ and $n_t$ are the amounts of actually collided nucleons from the projectile and target in the fixed-target mode~(they depend on the impact parameter); $m_N = 0.938$~GeV is the mass of nucleon; $\rho(0)$ is the density in the center of the fireball at the initial-time moment; $\alpha$ determines the distribution of density in the fireball, we take $\alpha = 0.25$. $B(x_1; x_2)$ in Eq.\,(\ref{eq::t_B}) is the beta function~(Euler integral of the~first kind). The quantity $t_B$ corresponds to the moment of turning the evolution of the fireball to the ballistic motion.

Application of~Eq.\,(\ref{eq::t_B}) to our case gives~$t_B = 0.61-0.85$~fm/$c$, depending on the impact parameter. The corresponding evolutions of the inverse Hubble parameter are shown in~Fig.\,\ref{fig::Hz} by the dashed colored lines for two values of the initial moment:~1.8~fm/$c$~(the first touch moment) and~4.07~fm/$c$~[estimate from~Eq.\,(\ref{eq::Hz_bs})]. Both sets of the curves representing~Eq.\,(\ref{eq::H_Bondorf}) go close to the data. 
On the simulated data for nucleons, the duration of growth of the Hubble parameter before the ballistic motion (the decreasing part of~$1/H_z$ after its second peak) is close to the prediction from~Eq.\,(\ref{eq::H_Bondorf}).

\subsection{Transverse expansion}
\label{ssec::Htransv}
\begin{figure}[h]
    \centering
    \includegraphics[width = \textwidth]{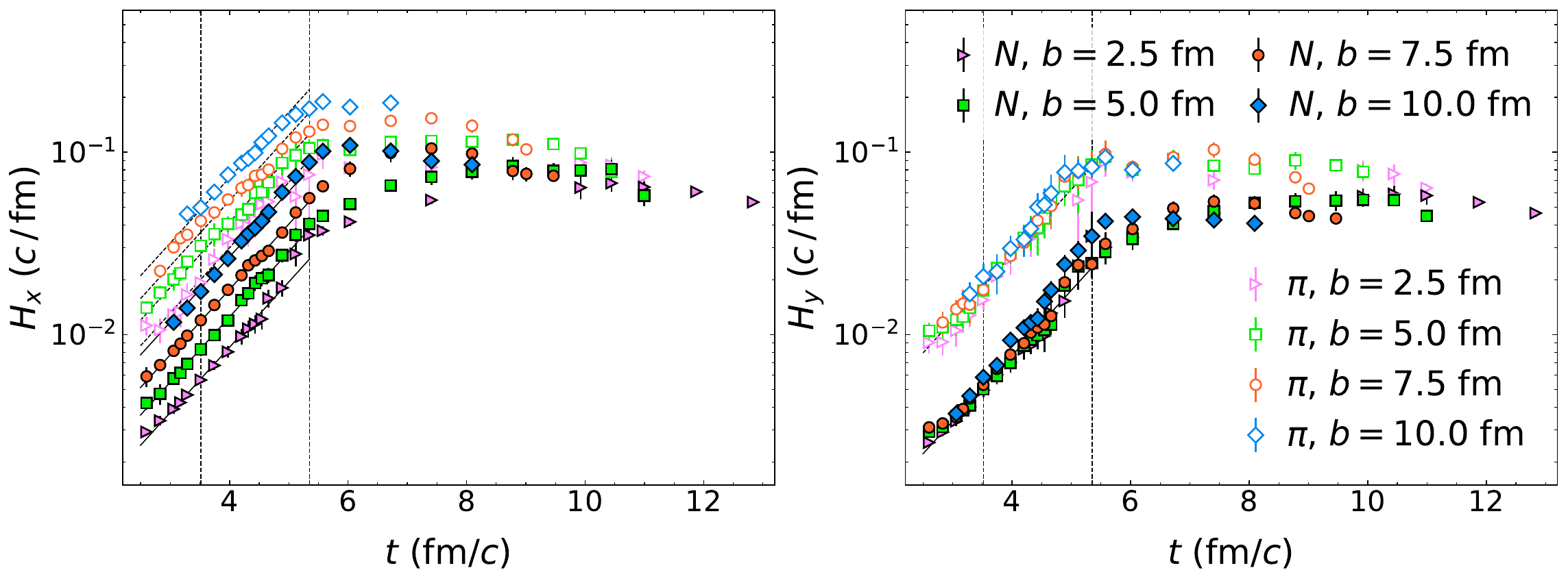}
    \caption{The Hubble parameter in the transverse directions. The slanted lines show exponential fits. Two vertical dashed lines are for the maximal overlapping moment of time and for the last touch moment.}
    \label{fig::Htransv}
\end{figure}

The expansion of the fireball in the transverse directions is qualitatively different from the longitudinal expansion. The Hubble parameters in both transverse directions are shown in~Fig.\,\ref{fig::Htransv}. The dependencies for pions and nucleons are well separated. The results in the~$y$ direction do not depend on the impact parameter. The data points for~$H_x$ are everywhere higher than the corresponding points for~$H_y$. This is in line with the statement~(see~Ref.\,\cite{Amelin:2007ic}) that a maximal transverse flow is along the~$x$ axis. One can observe a clear exponential growth of the Hubble parameter with time before the last touch moment. This resembles the cosmological inflation; however, the expansion rate in our case is much higher than in cosmology.

\begin{figure}[h]
    \centering
    \includegraphics[width = 0.45\textwidth]{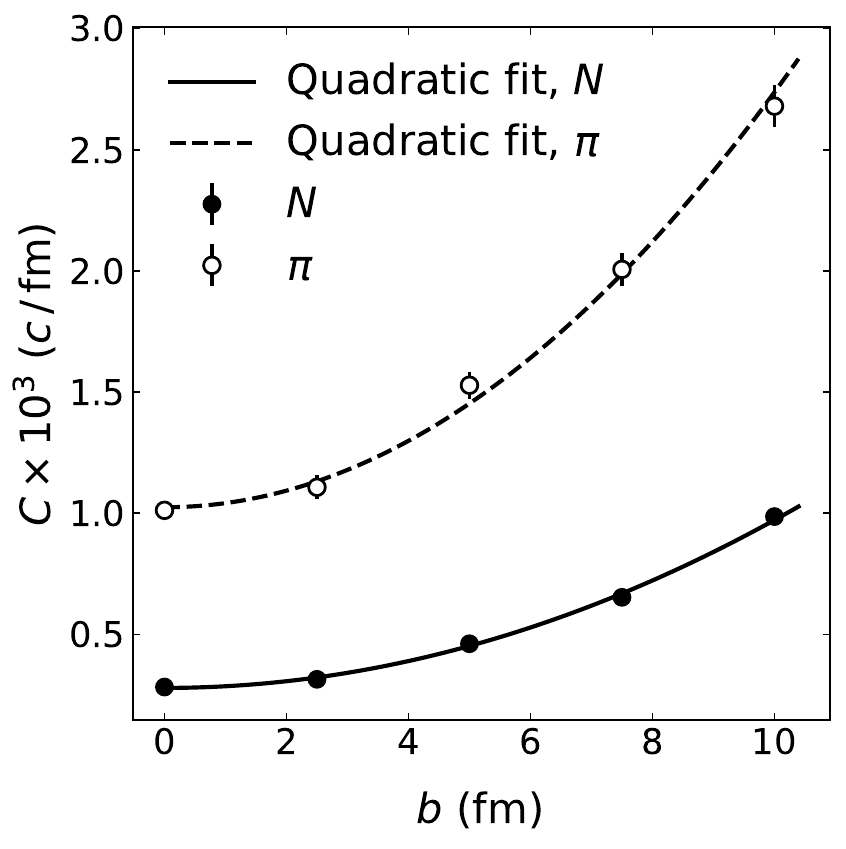}
    \caption{The scale factor~$C(b)$ from~Eq.\,(\ref{eq::exp}) for pions and nucleons.}
    \label{fig::ScaleN}
\end{figure}

For the parts with exponential growth, we perform an appropriate combined fit with a common power coefficient for the impact parameters and particle species.
In the~$x$ direction the dependences for each impact parameter are considered individually, while in the~$y$ direction they are averaged over the impact parameters separately for pions and nucleons. The estimated common power coefficient of the fit functions
\begin{equation}
    \label{eq::exp}
    H_i(b, t) = C(b)\,e^{k t}, \qquad i \in \{x,\, y\}
\end{equation}
is $k = 0.824(7)$~(fm/$c$)$^{-1}$. The scale parameters of~Eq.\,(\ref{eq::exp}) are shown in Fig.\,\ref{fig::ScaleN}. Here we suppose that $C(0)$ equals the scale factor obtained from the fits for~$H_y$, since for central collisions the~$x$ and~$y$ directions are equivalent and~$H_y$ is independent of the impact parameter~(see~Fig.\,\ref{fig::Htransv}). 
The obtained dependences of the scale factors on the impact parameter are well described by quadratic functions. 
This leads to the following parametrization of the ``microscopic'' Hubble parameter in the transverse directions:
\begin{equation}
    \label{eq::Htransv}
    H_x(b, t) = (a_0 + a_2 b^2) e^{k t}, \qquad H_y(t) = H_x(b = 0, t).
\end{equation}
The parameters of these expressions are summarized in Table~\ref{tab::Htransv}. 

\begin{table}[h]
\centering
\caption{Estimated parameters of Eq.\,(\ref{eq::Htransv}).}
\label{tab::Htransv}
\begin{tabular}{c|c|c|c}
\hline
& \rule{0pt}{1.1em}$a_0$, (fm/$c$)$^{-1}$, $\times 10^{-3}$ & $a_2$, $c$/fm$^3$, $\times 10^{-5}$ & $k$, (fm/$c$)$^{-1}$\\
\hline
Nucleons & 0.28(1) & 0.69(2) & \multirow{2}{*}{0.824(7)}\\
Pions    & 1.02(3) & 1.71(8) & \\
\hline
\end{tabular}
\end{table}

After separation of the nuclei, the obtained values of the ``microscopic'' Hubble parameter lie in the interval $H = 0.02 - 0.2$~(fm/$c$)$^{-1}$. This is consistent with the one given in~Ref.\,\cite{Baznat:2015eca}. Nevertheless, our interval is significantly wider. This difference mainly comes from the consideration of separate particle species and separate transverse directions in our calculations.

\begin{figure}[h]
    \centering
    \includegraphics[width = 0.95\textwidth]{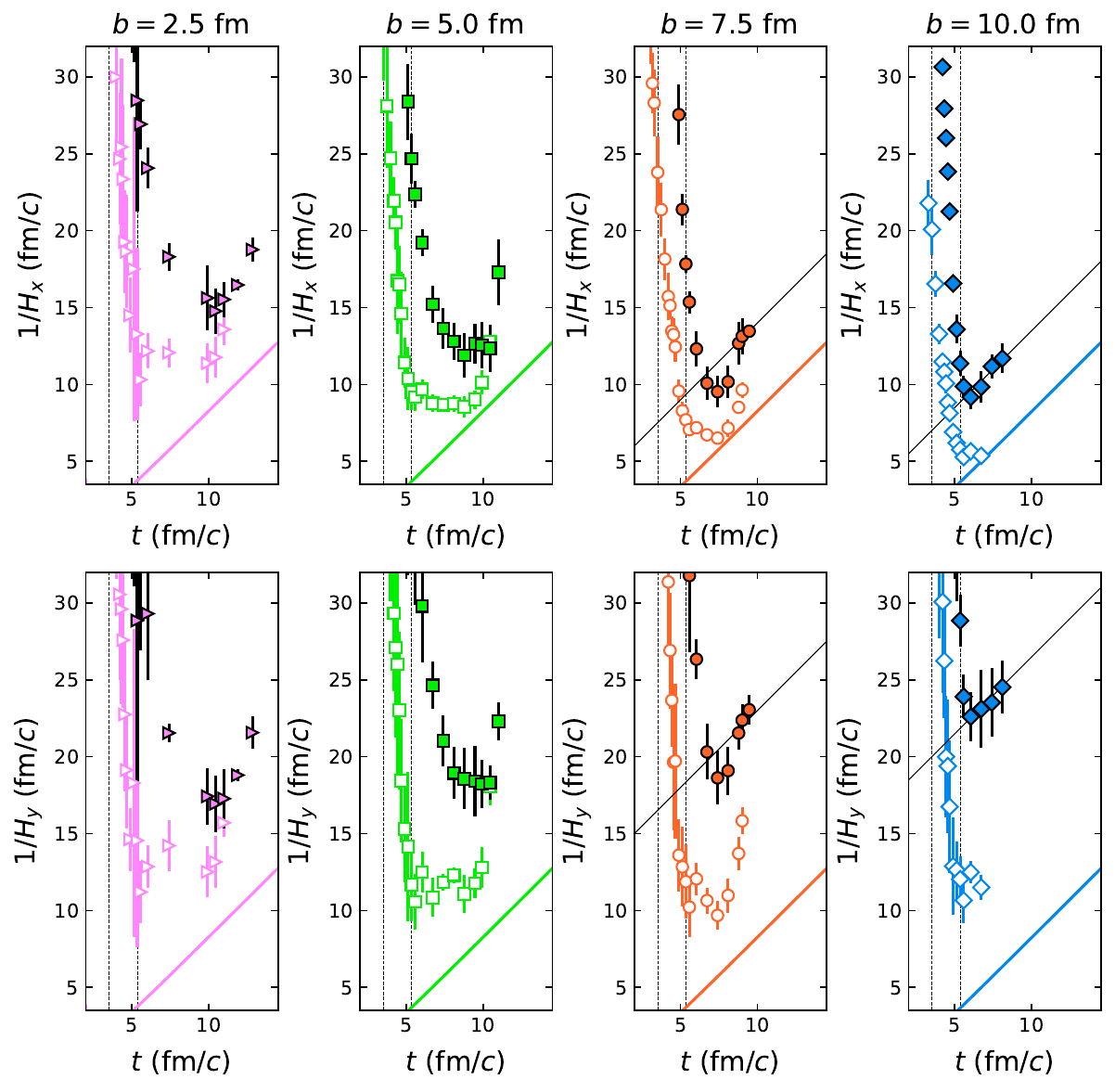}
    \caption{The late-time behavior of the ``microscopic'' Hubble constant. The slanted black lines show the ballistic-scattering lines drawn through the last points of the subplots. The colored lines represent the results from~Eq.\,(\ref{eq::H_Bondorf}). Two vertical dashed lines are for the maximal overlapping moment of time and for the last touch moment.}
    \label{fig::Htransv_linScale}
\end{figure}

Regarding high-time behavior of the ``microscopic'' Hubble parameter, it is natural to expect that at large enough time no forces act on the particles, so their motion is the ballistic one, and the evolution of the Hubble parameter is described by~Eq.\,(\ref{eq::Hz_bs}), where $C = 1$. Our data show that for the transverse directions we do not reach these times~(see Fig.\,\ref{fig::Htransv_linScale}). We can see just some hints of ballistic motion for nucleons at~$b \geq 7.5$~fm~(black slanted lines in~Fig.\,\ref{fig::Htransv_linScale}). However, a more detailed investigation of this time region is needed.

The statement that we probably do not reach freeze-out time is in line with the results of~Refs.\,\cite{Amelin:2006qe, Amelin:2007ic}. There, the estimated freeze-out time is much larger than $(t_{\text{LT}} - t_{\text{FT}}) \approx 0.3$~fm/$c$ at energies considered in that research.

\section{Conclusions}

The evolution of the ``microscopic'' Hubble parameter is obtained for~Au+Au collisions at~$\sqrt{s_{NN}} = 7.8$~GeV and the impact parameters~$b = 2.5$, 5.0, 7.5, and 10~fm separately for pions and nucleons. It is qualitatively different for the longitudinal and transverse directions.

For longitudinal expansion, ballistic motion with a shifted initial time moment is observed after separation of the residuals of the collided nuclei for both pions and nucleons at all impact parameters. This is in line with the previous investigations~\cite{Inghirami:2021zja, Tsegelnik:2022eoz}. The time shift is the same for pions and nucleons and is close to the moment of maximal particle yields. 

Before the separation of the nuclei, the evolution of the ``microscopic'' Hubble parameter is different for pions and nucleons and slightly depends on the impact parameter. The obtained values of the Hubble parameter here are in the interval $0.4 - 1.2$~(fm/$c$)$^{-1}$. For pions, the evolution can be described by the inverse cubic function. For nucleons, such parametrization is valid only up to~$t \approx 4$~fm/$c$. There is an indication that the evolution of the ``microscopic'' Hubble parameter starts from the same point~(the same time moment and the same value of~$H_z$) for both types of particles at all impact parameters. However, this statement should be checked in more details.

Our results for nucleons are compared with the solution of nonrelativistic equations of hydrodynamics for an isentropically expanding nucleon sphere~\cite{Bondorf:1978kz}. Both results are close to each other before the maximal overlapping moment
and after $t \approx 4.5$\,fm/$c$~[if one uses the estimate from~Eq.\,(\ref{eq::Hz_bs}) as the initial time moment in~Eq.\,(\ref{eq::H_Bondorf})].

For the transverse directions, almost until the last touch moment, the behavior of the ``microscopic'' Hubble parameter reflects a ``little inflation'', an analog of cosmological inflation but, in our case, with a much faster expansion rate. The Hubble parameter here grows exponentially with time. Some time after the separation of the nuclei, the Hubble parameter reaches its maximum~(both the position of the maximum and the value of the Hubble parameter there depend on the impact parameter and particle species, see~Fig.\,\ref{fig::Htransv_linScale}), and then starts to decrease. We do not observe the ballistic behavior in the transverse directions, just some hints of it at higher impact parameters. The obtained values of the Hubble parameter in the transverse directions lie in the interval~$0.02 - 0.2$~(fm/$c$)$^{-1}$ after the separation of the nuclei.

The curl of the velocity field
\begin{equation}
    \vec{v}_{\text{H}} = \sum_{i \in \{x, y, z\}} H_i\, x_i\, \vec{e}_i,
\end{equation}
where the Hubble-like motion takes place, equals zero, so it can be considered as at least part of the irrotational component  of the velocity field in its Helmholtz decomposition. The vorticity field plays the role of the solenoidal component in the decomposition.

It is interesting to compare the obtained results with the cosmological Hubble constant. 
The last is~$H_{\text{cosm}} = 70$~(km/s)/Mpc~$\sim 10^{-41}$~(fm/$c$)$^{-1}$, which is about~40 orders of magnitude smaller than the typical values obtained in~HIC. This correlates with a huge difference in values of ``micro'' and ``macro'' quantities, like acceleration~\cite{Prokhorov:2025vak} and vorticity~\cite{Vergeles:2022mqu}.

The results are obtained using data simulated within the PHSD transport model. The experimental determination of the ``microscopic'' Hubble constant is more complicated.

Explicit determination of the Hubble parameter at a given time moment requires simultaneous knowledge of the positions and velocities of particles inside the fireball. 
It seems possible to estimate the value of the ``microscopic'' Hubble parameter at freeze-out using the blast-wave fit of invariant $p_T$ spectra to get the velocity of the particles at freeze-out together with femtoscopy data on the size of the fireball~\cite{Nedorezov:2025sya, Kisiel:2025jbg, Fabbietti:2020bfg}.

A more involved possibility is to study an analog of the cosmic microwave background in HIC~\cite{Mocsy:2011xx, Heinz:2013wva}. The ``microscopic'' Hubble parameter should be imprinted into its structure, as it is in the case of cosmology.

Also, some indirect study of the Hubble-like motion seems possible. The lack of simultaneous knowledge of the coordinates and velocities of particles at any time moment can be circumvented to some extent by connecting the Hubble parameter with coordinate- and velocity-dependent phenomena, such as polarization~\cite{Vitiuk:2019rfv} and anisotropic flows~\cite{Lacey:2013qua}. The anisotropy of the ``microscopic'' Hubble parameter described in Sec.~\ref{ssec::Htransv} should be related to anisotropic flows and polarization~\cite{STAR:2023eck} in some way. 

We expect Hubble's law and possibly ``little inflation'' to manifest themselves at higher and lower energies. This requires additional investigations.

\section*{Acknowledgments}
The authors are grateful to  E.\,E.~Kolomeitsev, N.\,S.~Tsegelnik, R.~Lednicky, V.\,I.~Kolesnikov and V.\,A.~Plotnikov for fruitful discussions and useful remarks.
We would like to thank HybriLIT group of JINR for providing computational resources at HybriLIT and Govorun clusters.

\bibliographystyle{apsrev1}
\bibliography{HubblePaper}

\end{document}